\documentclass[prl,twocolumn,superscriptaddress,showpacs]{revtex4}
\usepackage{epsfig,dcolumn,amsmath,latexsym,graphicx}
\begin{document}
\newcommand{\IV}{~{\it I-V$_b$}~} \newcommand{\eV}{\mbox{ eV}}
\newcommand{\meV}{\mbox{ meV}} \newcommand{\Vswitch}{V_{S}}
\newcommand{\V}{\mbox{V}} \newcommand{\HOMO}{\mbox{HOMO}}
\newcommand{\LUMO}{\mbox{LUMO}} \newcommand{\Siemen}{\mbox{S}}
\newcommand{\degrees}{$^\circ$}

\title{Conductance switching in a molecular device: the role of
sidegroups and intermolecular interactions} \author{Jeremy Taylor,
Mads Brandbyge, and Kurt Stokbro} \affiliation{Mikroelektronik Centret
(MIC), Technical University of Denmark (DTU), Building 345 east,
DK-2800 Kongens Lyngby, Denmark}
\begin{abstract}
We report first-principles studies of electronic transport in
monolayers of Tour wires functionalized with different side groups.
An analysis of the scattering states and transmission eigenchannels
suggests that the functionalization does not strongly affect the
resonances responsible for current flow through the
monolayer. However, functionalization has a significant effect on the
interactions within the monolayer, so that monolayers with NO$_2$ side
groups exhibit local minima associated with twisted conformations of
the molecules.  We use our results to interpret observations of
negative differential resistance and molecular memory in monolayers of
NO$_2$ functionalized molecules in terms of a twisting of the central
ring induced by an applied bias potential.
\end{abstract}
\maketitle

An important goal in the study of molecular electronics is to identify
molecules that can be combined to perform logical
functions\cite{JoGiAv00}.  In particular, Reed and
co-workers~\cite{ChReRaTo99,ChWaReRaPrTo00,Chen-2002-CHEMPHYS,ReChRaPrTo01}
have studied the electrical properties of a set of phenyl-ethylene
oligomers (known as {\it Tour wires } (TW)\cite{ToKoSe98})
functionalized with different side groups, and demonstrated that such
molecules can show negative differential resistance
(NDR)\cite{ChReRaTo99,ChWaReRaPrTo00} and a ''molecular
memory''\cite{Chen-2002-CHEMPHYS,ReChRaPrTo01} effect, in which case
molecules can be switched from a low conductance state to a high
conductance state by application of a voltage pulse.
A fundamental understanding of the microscopic mechanisms governing
the NDR and memory effects in these systems is still lacking.

Complementary studies, by Weiss and co-workers
\cite{DoMaKeBuMoStPrRaAlToWe01}, of isolated or small bundles of TWs
embedded in a monolayer of spacer alkyl molecules, demonstrated {\em
spontaneous} conductance switching. While Reed and co-workers only
observed switching effects for TWs with NO$_2$ sidegroups, Weiss
et. al.  reported spontaneous switching of all the molecules in their
study but had limited success in {\em inducing} switching by using a
voltage pulse\cite{DoMaKeBuMoStPrRaAlToWe01}. Theoretical studies of
properties of isolated TWs have emphasized the role of charging of the
molecule and subsequent localization/delocalization of molecular
orbitals \cite{SeZaTo00,SeZaDe01} or bias-induced alignment of
molecular orbitals on the first and last phenyl rings of twisted
TWs\cite{Cornil-2002-JACS}.

\begin{figure}
\begin{center}
\leavevmode \epsfxsize=84mm
\includegraphics[width=\columnwidth]{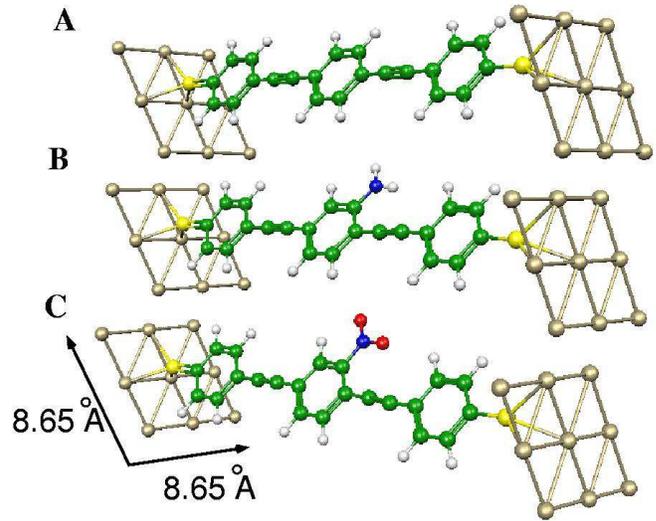}
\end{center}
\caption{
\protect\small Geometry of monolayers A, B, and C connected with two
Au (111) surfaces. Color codes: C(green), H(white), O(red), N(blue),
S(yellow), Au(gold).  }
\end{figure}

In this paper we use the TranSiesta package\cite{Brandbyge-2002-PRB}
 to investigate the effect of different sidegroups on both the
 electrical and structural properties of TW's in self assembled
 monolayers, and in this way gain new insight into the microscopic
 origin of the observed NDR and memory effects. TranSIESTA is a
 non-equilibrium Greens function based electronic transport program
 founded on density functional theory. It treats the entire system
 selfconsistently under finite bias conditions, and the only major
 approximation in the method is the choice of exchange correlation(XC)
 functional. \cite{tech3} We have previously shown that electrical
 properties of different metallic wires calculated using the method
 are in excellent agreement with experimental
 data\cite{Nielsen-2002-PRL}.

The nanopore structures in which NDR and memory effects have been
observed are formed by evaporating the molecules onto a small Au
surface inside a Si$_3$N$_4$ structure, and then evaporating another
Au electrode onto the molecular monolayer\cite{ChReRaTo99}.  Since
there is no detailed information about the structure of the monolayer,
our theoretical analysis of the problem proceeds by forming an
idealized model of the electrode/monolayer/electrode system and
comparing electron transport in three different molecules\cite{tech1}.
We use an Au(111) surface in a 3x3 unit cell for the electrodes and
assume the TWs are chemisorbed to the surfaces through strong thiol
bonds\cite{bonding}, as illustrated in Fig. 1.  monolayer A consists
of a bare TW with thiol end groups, while monolayers B and C have been
functionalized with NH$_2$ and NO$_2$ side groups, respectively.

We first investigate the effect of the sidegroups on the electrical
properties of the monolayers. The current through the monolayers is
determined by the quantum mechanical probability for electrons to
tunnel from one electrode to the other, and is calculated using the
Landauer formula\cite{Da96}: $I = \int_{\mu_L}^{\mu_R} T(E,V_b) dE $,
where $\mu_{L/R} = \pm e V_b $ and $T(E, V_b)$ is the transmission
probability for electrons incident at an energy $E$ through a device
under a potential bias $V_b$ ({\em n.b} $V_b > 0 $ corresponds to hole
injection from the right electrode).  The general shape of the
zero-bias tranmsission spectra, shown as $T(E,V_b=0 V)$ in Fig. 2a, is
similar for all three systems.  The zero bias conductance $G =
\frac{e^2}{h} T(E=\mu_{l/r},V_b=0) $, given by $2.0$ $ \mu\Siemen$,
$2.3$ $\mu\Siemen$, $1.9$ $ \mu\Siemen$, for monolayers A, B, and C,
is dominated by the tail of a broad highest-occupied-molecular-orbital
(HOMO) resonance (at $E_{\HOMO}=-0.84$, $-0.65$, and $-0.85 \eV$ for
monolayers A, B, and C).  The narrower
lowest-unoccupied-molecular-orbital (LUMO) resonance (at
$E_{\LUMO}=1.56$ , $1.59$, and $1.14 \eV$ for A, B, and C
respectively) contributes less to the conductance and thus holes are
the dominant carriers.  The NH$_2$ group is electron donating, so that
$E_{\HOMO,B} - E_{\HOMO,A} \approx 200 \meV$ while the NO$_2$ group is
electron accepting so $E_{\LUMO,C} - E_{\LUMO,A} \approx -400 \meV$.

\begin{figure}
\begin{center}
\leavevmode \epsfxsize=84mm
\includegraphics[width=\columnwidth]{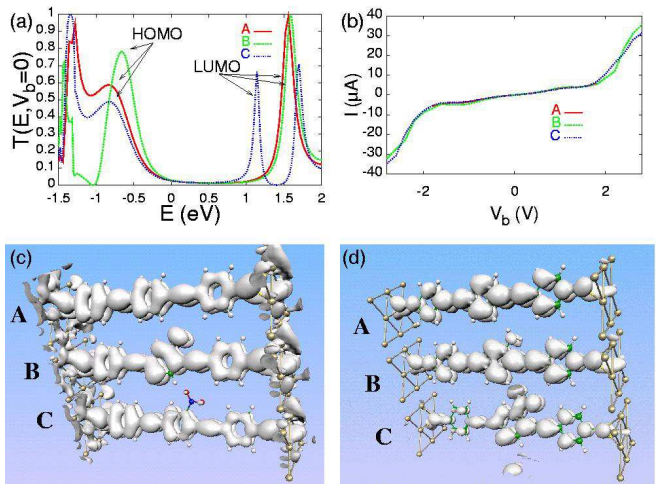}
\end{center}
\caption{
\protect\small (a) Zero-bias tranmssion $T(E,V_b=0 V)$ versus incident
$E$ for monolayers A, B, and C. HOMO and LUMO resonances are
indicated.  (b) \IV characteristics for monolayers A, B, and C (c)
Isosurface of transmission eigenchannel at the HOMO resonance for
monolayers A, B, and C ($E=-0.84$, $-0.65$, and $-0.85 \eV$
respectively) (d) Isosurface of transmission eigenchannel at the LUMO
resonance for monolayers A, B, and C ($E=1.56$ , $1.59$, and $1.14
\eV$ respectively) }
\end{figure}

To calculate the \IV spectrum we performed selfconsistent calculations
for biases in the range -2.8 V to 2.8 V in steps of 0.2 V.  We note
that the charge on the molecule is {\em not}
fixed\cite{Brandbyge-2002-PRB} and adjusts itself to minimize the free
energy\cite{Todorov-2000-PHYSMAGB} as the left/right electrochemical
potentials are changed. For all three systems we find that the
molecules are close to charge neutrality, and the charge on the
molecule $Q_M$ , as determined by a Mullliken population analysis,
change by less than 0.05 e as the bias voltage $V_b$ is varied. The
\IV spectra, shown in Fig. 2b, are very similar, increasing slowly at
first but then increasing rapidly around $ 2 V$ where the resonances
come into alignment with the bias window.  The main effect of $V_b$ on
$T(E,V_b)$ is to sample more and more of the resonance while resonant
peak height and position vary slowly with $V_b$\cite{Taylor-2002-PRL}.

When the molecules interact with the Au(111) electrodes, the molecular
levels broaden into a continuum\cite{Da96}.  The eigenstates of the
whole metal/monolayer/metal system consist of scattering
states\cite{EmKi98,Taylor-2000-PHD}, which are molecular orbital-like
in the molecule, and Bloch wave-like in the metal slabs.  If an
orbital is delocalized across the molecule, an electron that enters
the molecule at the energy of the orbital has a high probability of
reaching the other end and thus there is a corresponding peak in the
transmission probability $T(E,V_b)$, as illustrated in Fig. 2a.  By
calculating the continuum eigenstates at the resonance energies
($E_{\HOMO}, E_{\LUMO}$), the orbitals that are responsible for
current flow through the molecule can be analyzed.  The HOMO
resonances, illustrated in Fig. 2c, resemble the HOMO of the isolated
molecules while the narrower LUMO resonances, illustrated in Fig. 2d,
resemble the LUMO of the isolated molecules.
There are minor differences between the orbitals: the NH$_2$ side
group participates in the HOMO resonance but not in the LUMO
resonance, while the NO$_2$ group participates in the LUMO resonance
and not the HOMO resonance, consistent with their respective
donating/accepting characters.  It is, however, clear from both the
shape of the transmission curves in Fig. 2a and the orbitals in
Figs. 2c and 2d that the resonant transmission peaks in monolayers A,
B, and C are related to the delocalized nature of the $\pi$-orbitals
of the bare TWs and is not strongly affected by the functionalization.


\begin{figure}
\begin{center}
\leavevmode \epsfxsize=84mm
\includegraphics[width=\columnwidth]{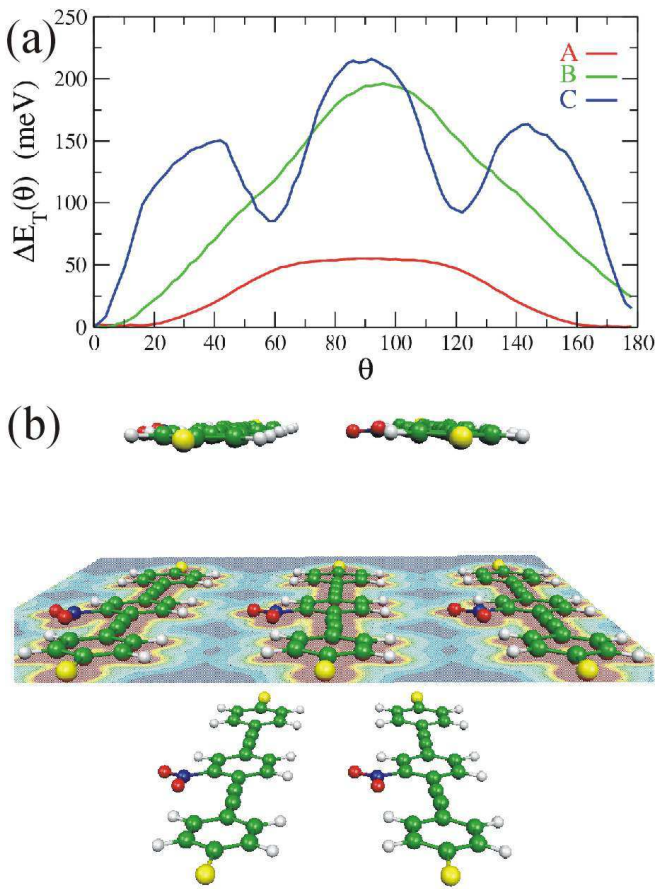}
\end{center}
\caption{
\protect\small (a) Energy versus rotation angle for Molecules A, B,
and C in Au(111) 3x3 unit cell. The energy is calculated within the
Perdew, Burke and Ernzerhof (PBE) approximation for the
exchange-correlation functional\cite{Perdew-1996-PRL}. (b) Contourplot
of the effective potential between TW's with NO$_2$ sidegroups. Note
the bond formation between the O atom and the H atom on the
neighboring TW.}
\end{figure}

The energetics of the monolayers are, however, strongly affected by
the functionalization.  The triple bond between the phenyl rings is
rotationally symmetric and thus constitutes an easy axis about which
the middle ring can rotate. The differences in total energy as a
function of the rotation angle of the middle ring, $\theta$, for the
molecules arranged in the 3x3 unit cell of Au(111) are shown in
Fig. 3a.
The intermolecular interaction energy is strongly dependent on the
functionalization: there is a cost of $60 \meV$ ($180 \meV$) to rotate
the ring 90\degrees in monolayer A (B), and the flat molecule is
clearly favoured.  Monolayer C shows a quite different behaviour, it
acquires a local minimum at $\theta = 60$\degrees and $
120$\degrees. To understand the origin of these minima we have plotted
the effective potential within the monolayer in Fig. 3b.  The
effective potential reveals the formation of a hydrogen bond between
the NO$_2$ group and a hydrogen atom on the neighboring TW. Similar
hydrogen bond formation involving NO$_2$ groups has also been observed
in other molecular layers\cite{Bohringer-2002-PRL}. From Fig. 3b it is
also clear that the minima at $60$\degrees and $ 120$\degrees can be
attributed to bond formation with a new neighboring molecule, and
these particular angles arise from the symmetry of the monolayer.

There is a barrier for rotation of the middle ring from $\theta = 0$
to $\theta = 60$\degrees of $\sim 150 \meV$, while the barrier from $
60$\degrees to $0$\degrees is $\sim 65 \meV$ We postulate that, as a
voltage pulse is applied, these barriers may change, and a transition
to a new molecular conformation with the middle ring twisted may be
induced.  The local minima only exist when the molecules are
functionalized with an NO$_2$ side group and thus such transitions
only exist when the molecules are functionalized with an NO$_2$ side
group \cite{Chen-2002-CHEMPHYS}.

Again because of the lack of detailed information about the monolayer,
we have selected a candidate conformation, which we designate
monolayer C@60, formed by twisting the middle ring of molecules in
monolayer C 60\degrees, for possible stabilization under applied bias.
The transmission spectra for monolayer C@60 is shown in Fig. 4a and
its \IV characteristics in Fig. 4d. The conductance $G\approx .12
\mu\Siemen $ is roughly 16 times smaller than monolayer C while the
current at $V_b = 2 V$ is $\approx 5$ times smaller. To lowest order,
$G$ is proportional to the product of the matrix elements between
$\pi$-orbitals on neighbouring phenyl rings so that a rotation of the
middle ring will reduce $G$ by a factor $\sim \cos^4(\theta) =
\frac{1}{8}$, in rough agreement with the calculated values.

The transmission eigenstates for electrons travelling from left to
right at $E_{\HOMO} = -0.85 \eV$ and $E_{\LUMO} = 0.99 \eV$ are
illustrated in Fig. 4b.  The HOMO resonance is localized on the phenyl
ring nearest the left electrode while the LUMO state corresponds to an
orbital of the middle ring, weakly coupled to the first and last
rings.

To investigate whether monolayer C@60 could be stabilized with respect
to monolayer C as the bias voltage is increased, we show in Fig. 4b
the Kohn-Sham total energy of monolayers C and C@60. At $V_b = 0 $,
monolayer C is lower in energy than monolayer C@60. However, when the
bias is increased above 2 V, monolayer C@60 becomes lower in energy,
leading to an expected conformational change.  While there are issues
surrounding the evaluation of total energies for such systems
\cite{Todorov-2000-PHYSMAGB,energy-note}, Fig. 4b serves to
illustrate that different conformations may be stablized with respect
to each other by an external bias.  In this case, the calculation
suggests a transformation of monolayer C to monolayer C@60 at around $
2 V $.  The exact value of the switching voltage will depend on the
coverage, detailed structure and size of the monolayer, as well as how
the potential drops across the molecules, which can be affected by
electrode coupling\cite{Taylor-2002-PRL}.

\begin{figure}
\begin{center}
\leavevmode \epsfxsize=84mm
\includegraphics[width=\columnwidth]{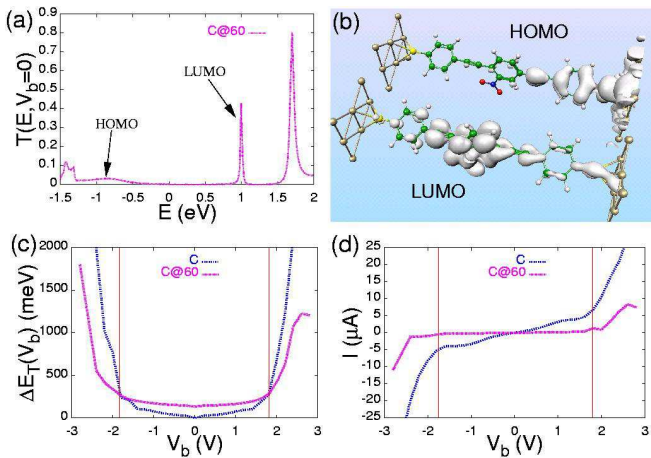}
\end{center}
\caption{\protect\small
(a) Zero-bias transmission $T(E,V_b=0)$ for monolayer C@60.  (b)
Transmission eigenchannels corresponding to HOMO and LUMO resonances
of C@60. (c) Total energy of monolayer C(blue) and monolayer
C@60(magenta) as a function of bias potential.  (d) \IV
characteristics of monolayer C(blue) and monolayer C@60(magenta).}
\end{figure}

The local minima in Fig.~3a are stabilized by the NO$_2$ sidegroup,
which correlates well with that the molecular memory effect is only
observed in monolayers with NO$_2$ side
groups\cite{Chen-2002-CHEMPHYS}. Furthermore, the energy barrier for
rotation of $65-150 \meV$ derived from our crude model is in rough
agreement with the $80 \meV$ barrier height extracted from bit
retention times\cite{Chen-2002-CHEMPHYS}, thus conformational changes
stabilized by intermolecular interactions is a good candidate for
explaining such phenomena.

Because the minima emerge from interactions within the monolayers, the
above effects may only be observed in nanopore experiments
\cite{ChReRaTo99,ChWaReRaPrTo00,Chen-2002-CHEMPHYS,ReChRaPrTo01} and
not in experiments of single molecules. In the STM experiments of
Weiss and co-workers\cite{DoMaKeBuMoStPrRaAlToWe01} on single or small
bundles of molecules, the conductance switching may be related to
different conformations formed by steric interactions between the TWs
and the insulating alkyl molecules. This could explain why the
observed switching behaviour was independent of the functionalization
of the molecules.

In conclusion, we find that functionalization of TWs has a stronger
effect on the energetics of the monolayers than on the orbitals
responsible for current transport and a better understanding of the
intermolecular interactions in such monolayers could hopefully be
exploited in order to design molecular electronic devices with
specific properties.


{\bf Acknowledgements:} We would like to thank J.-L. Mozos and
P. Ordejon for help in implementing {\em TranSIESTA}, and J. Cornil
for valuable discussions. We acknowledge funding from the SNF (M.B.),
STVF (K.S.,J.T.), NSERC (J.T.), EU SANEME IST-1999-10323, Direkt\o r
Ib Henriksens fond, and Danish Center for Scientific Computing (DCSC).


\begin{thebibliography}{26}
\bibitem[{tec({\natexlab{b}})}]{tech1}
\bibinfo{note}{We use a Double $\zeta$+Polarization basis set for the organic
  molecule, and a double $\zeta$ for the gold $s$ channel and single $\zeta$
  for the gold $p$ and $d$ channel.}
\bibitem[{bon()}]{bonding}
\bibinfo{note}{The molecular geometry is obtained by first optimizing the
  geometry of the free molecule with H atoms attached to the Sulfur atoms, then
  placing the molecule at the FCC positions with an Au-S distance of 2.45 \AA
  \cite{AnCuGr00}, and finally relaxing the molecule in the sandwich structure
  with fixed Au atoms.}
\bibitem[{ene()}]{energy-note}
\bibinfo{note}{The statistically most likely geometry is that which minimizes
  the free energy: $ E - \mu_L Q_L - \mu_R Q_R $ where $Q_L$ and $Q_R$ are the
  charges injected from the left and right
  electrodes\cite{Todorov-2000-PHYSMAGB}. The quantites $Q_L$ and $Q_R$ are,
  for technical reasons, very expensive to calculate. Furthermore, in order to
  conclusively determine which geometry has the lowest free energy, we should
  allow for relaxations of the atomic coordinates under the external bias.}
\end{thebibliography}
\end{document}